\documentclass[prb, onecolumn,preprint]{revtex4}
\usepackage{graphicx}
\usepackage{latexsym}
\usepackage{amsmath}
\usepackage{amssymb}
\usepackage{epstopdf}
\usepackage{dcolumn}
\usepackage{setspace}  
\begin{document}
\title{The underdoped cuprates as fractionalized Fermi liquids: transition to superconductivity}

\author{Eun Gook Moon and Subir Sachdev}
\affiliation{Department of Physics, Harvard University, Cambridge MA 02138}

\date{\today}

\begin{abstract}
We model the underdoped cuprates using fermions moving in a background with local antiferromagnetic
order. The antiferromagnetic order fluctuates in orientation, but not in magnitude, so that there is no
long-range antiferromagnetism, but a `topological' order survives. 
The normal state is described as a fractionalized Fermi liquid (FL*),
with electron-like quasiparticles coupled to the fractionalized excitations of the fluctuating antiferromagnet.
The electronic quasiparticles reside near
pocket Fermi surfaces enclosing total area $x$ (the dopant density), centered 
away from the magnetic Brillouin zone
boundary. The violation of the conventional Luttinger theorem is linked to a `species doubling' of these quasiparticles.
We describe phenomenological theories of the pairing of these quasiparticles, and show that a large class 
of mean-field theories generically 
displays a nodal-anti-nodal `dichotomy': the interplay of local antiferromagnetism and pairing leads to 
a small gap near the nodes of the $d$-wave pairing along
the Brillouin zone diagonals, and a large gap in the anti-nodal region. 
\end{abstract}

\maketitle

\section{Introduction}

The nature of the ground state in the underdoped regime of the hole-doped cuprate superconductors
remains a central open issue. 
Angle resolved photoemision spectroscopy (ARPES) and scanning tunneling microscopy (STM) have been the main tools to explore such a regime. In both probes,
an unexpected angular dependence of the electron spectral gap function has been revealed: 
a `dichotomy' between the nodal and anti-nodal regions of the Brillouin zone in the superconducting 
state \cite{boyer,dichot0,dichot1,dichot2,dichot3}. Specifically, this dichotomy is realized by 
 deviations in the angular dependence of the gap from that of a short-range
$d$-wave pairing amplitude $\sim (\cos k_x - \cos k_y)$.

This paper will describe the superconducting instabilities of a recently developed model \cite{pockets} of the 
normal state of the 
underdoped cuprates based upon a theory of fluctuating local antiferromagnetic 
order \cite{ssss,rkk1,rkk2,su2}. A related
normal state model of fluctuating antiferromagnets has
been discussed by Khodas and Tsvelik \cite{tsvelik1}, who obtained results on the influence of spin-wave fluctuations
about the ordered state similar to ours \cite{pockets}. These results have been found to agree well with ARPES
observations \cite{tsvelik2,peter3,Meng,Damascelli}. Another approach using fluctuating antiferromagnetism to model the
underdoped cuprates has been discussed recently 
by Sedrakyan and Chubukov \cite{andrey}. We will also connect with the scenario emerging from recent
dynamical mean-field theory (DMFT) studies \cite{dmft1,dmft2,sordi}.

The theory of Ref.~\onlinecite{pockets} describes the normal state in the underdoped regime 
as a fractionalized Fermi liquid (FFL or FL*), although this identification was not explicitly made in that paper.
So we begin our discussion by describing the the structure of the FL* phase. 

The FL* phase is most naturally
constructed \cite{ffl1,ffl2} using a Kondo lattice model describing a band of conduction electrons coupled to 
lattice of localized spins arising from a half-filled $d$ (or $f$) band.  
The key characteristics of the FL* are ({\em i}\/) a `small' Fermi surface whose volume is determined
by the density of conduction electrons alone, and ({\em ii\/}) the presence of gauge and fractionalized 
neutral spinon excitations of a spin liquid. In the simplest picture, the FL* can be viewed in terms
of two nearly decoupled components, a small Fermi surface of conduction electrons and 
a spin liquid of the half-filled $d$ band. The FL* should be contrasted from the 
conventional Fermi liquid, in which there is a `large' Fermi surface
whose volume counts both the conduction and $d$ electrons: such a heavy Fermi liquid phase 
has been observed in many `heavy fermion' rare-earth intermetallics.
Recent experiments on YbRh$_2$(Si$_{0.95}$Ge$_{0.05}$)$_2$ have 
presented evidence \cite{paschen} for an unconventional phase, 
which could possibly be a FL*.

A concept related to the FL* is that of a 
``orbital-selective Mott transition'' \cite{osmt} (OSMT), as discussed in the
review by Vojta \cite{vojta}. For latter, we begin with a multi-band model, like the lattice Anderson
model of conduction and $d$ electrons, and have a Mott transition to an insulating state on only a subset of the bands
(such as the $d$ band in the Anderson model). The OSMT has been described so far using
dynamical mean field theory (DMFT), which has an over-simplified treatment of the Mott insulator.
In finite dimensions, any such Mott insulator must not break lattice symmetries which increase the size
of a unit cell, for otherwise the state reached by the OSMT is indistinguishable from a conventionally
ordered state. Thus the Mott insulator must be realized as a fractionalized spin liquid with collective
gauge excitations; such gauge excitations are not present in the DMFT treatment. 
With a Mott insulating spin liquid, the phase reached by the OSMT becomes a FL*.

Returning our discussion to the cuprates, there is  
strong ARPES evidence for only a single band of electrons, with a conventional Luttinger volume
of $1+x$ holes at optimal doping and higher (here $x$ is the density of holes doped into the half-filled insulator).
Consequently, the idea of an OSMT does not seem directly applicable.
However, Ferrero {\em et al.} \cite{dmft2} argued that an OSMT could occur in momentum space within
the context of a single-band model. They separated the Brillouin zone
into the `nodal' and `anti-nodal' regions, and represented the physics using a 2-site DMFT solution.
Then in the underdoped region,  
the anti-nodal region underwent a Mott transition into an insulator, while the nodal regions
remained metallic. A similar transition was seen by Sordi {\em et al.\/} in studies with a 4-site 
cluster\cite{sordi}.
While these works offers useful hints on the structure of the intermediate energy physics,
ultimately the DMFT method does not allow full characterization of the different low energy quasiparticles
or the nature of any collective gauge excitations.

We turn then to the work of  Ref.~\onlinecite{pockets}, who considered a single band model of a fluctuating antiferromagnet.
Their results amount to a demonstration that a FL* state can be constructed also in a single band model,
and this FL* state will form the basis of the analysis of the present paper.
The basic idea is that the large Fermi surface is broken apart into pockets by local antiferromagnetic N\'eel order.
We allow quantum fluctuations in the 
orientations of the 
N\'eel order so that there is no global, long-range N\'eel order. However, spacetime `hedgehog'
defects in the N\'eel order are suppressed, so that 
a spin liquid with bosonic spinons and a U(1) gauge-boson excitation is 
realized \cite{mv,senthil}. 
Alternatively, the N\'eel order could develop spiral spin correlations, and suppressing $Z_2$ vortices in the spiral order
realizes a $Z_2$ spin liquid with bosonic spinons\cite{rs2,rsl}.
The Fermi pockets also fractionalize
in this process, and we are left with Fermi pockets of spinless fermions; the resulting phase was
called the algebraic charge liquid \cite{rkk1,rkk2,su2} (ACL). Depending upon the nature of the gauge excitations of 
the spin liquid,
the ACL can have different varieties: the U(1)-ACL and SU(2)-ACL were described in Refs.~\cite{su2},
and $Z_2$-ACL descends from these by a Higgs transition involving a scalar with U(1) charge 2,
as in the insulator \cite{rs2,rsl}.

Although these ACLs are potentially stable phases of matter, they are generically susceptible to
transformation into FL* phases.
As was already noted in
Ref.~\onlinecite{rkk1}, there is a strong tendency for the spinless fermions to found bound states
with the bosonic spinons, leading to pocket Fermi surfaces of quasiparticles of spin $S=1/2$ and charge
$\pm e$. Also, as we will review below, there is a `species-doubling' of these bound states \cite{wenholon,ssss,rkk1},
and this is crucial in issues related to the Luttinger theorem, and to our description of the superconducting
state in the present paper.
When the binding of spinless fermions to spinons
is carried to completion, so that Fermi surfaces of spinless fermions has been completely depleted,
we are left with Fermi pockets of electron/hole-like quasiparticles which enclose a total
volume of precisely $x$ holes \cite{pockets}. The resulting phase then has all the key characteristics
of the FL* noted above, and so we identify it here as a FL*. 
The U(1)-ACL and $Z_2$-ACL above lead to the conducting 
U(1)-FL* and $Z_2$-FL* states respectively.
Ref.~\onlinecite{pockets} presented
a phenomenological Hamiltonian to describe the band structure of these FL* phases.
Thus this is an explicit route to the appearance of an OSMT in a single-band, doped
antiferromagnet: it is the local antiferromagnetic order which differentiates regions of the Brillouin zone,
and then drives a Mott transition into a spin liquid state, leaving behind Fermi pockets of holes/electrons
with a total volume of $x$ holes.

We should note here that the U(1)-FL* state with a U(1) spin liquid is ultimately unstable
to the appearance of valence bond solid (VBS) order at long scales \cite{rs1}. However the $Z_2$-FL* 
is expected to describe a stable quantum ground state. The analysis of the fermion spectrum below remains the same for the two cases.

Phases closely related to the U(1)-FL* and $Z_2$-FL* appeared already in the work of Ref.~\onlinecite{ssss}.
This paper examined `quantum disordered' phases of the Shraiman-Siggia model \cite{ss}, 
and found states with small Fermi pockets, but no long-range antiferromagnetic order. The antiferromagnetic correlations where either collinear or spiral, corresponding to the U(1) and $Z_2$ cases. However, the topological 
order in the sector with neutral spinful excitations was not recognized in this work: these spin excitations
were described in terms
of a O(3) vector, rather than SU(2) spinor description we shall use here. Indeed, the topological order is required
in such phases, and is closely linked to the deviation from the traditional volume of the Fermi surfaces. \cite{ffl1,ffl2}

We also note another approach to the description of a FL* state in a single band model, in the work of 
Ribeiro, Wen, and Ran \cite{RW1,RW2,RW3}. They obtain a small Fermi surface of electron-like ``dopons'' 
moving in spin-liquid background. However, unlike our approach with gapped bosonic spinons (and associated connections
with magnetically ordered phases), their spinons are fermionic and have gapless Dirac excitation spectra centered
at $(\pm \pi/2, \pm \pi/2)$.

We will take the U(1)-FL* or $Z_2$-FL* 
state with bosonic spinon spin liquid as our model for the underdoped cuprates in the present paper.
We will investigate its pairing properties using a simple phenomenological model of $d$-wave
pairing. Our strategy will be to use the simplest possible model with nearest-neighbor pairing
with a $d$-wave structure, constrained by the requirement that the full square lattice translational
symmetry and spin-rotation symmetry be preserved. Even within this simple context, we will find
that our mean-field theories of the FL* state allows us to easily obtain the `dichotomy' in the pairing amplitude over a very broad range of parameters. We also note that the pocket Fermi surfaces of the FL* state will exhibit quantum oscillations in an applied
magnetic field with a Zeeman splitting of free spins, and this may be relevant to recent observations \cite{ramshaw}.

We mention here our previous work \cite{gs,superACL,QCpairing} on pairing in the parent ACL phase. These papers
considered pairing of spinless fermions, while the spin sector was fully gapped: this therefore
led to an exotic superconductor in which the Bogoliubov quasiparticles did not carry spin. 
In contrast, our analysis here will
be on the pairing instability of the FL* state, where we assume that the fermions have already bound into
electron-like quasiparticles, as discussed above and in more detail in Ref.~\onlinecite{pockets}.
The resulting Bogoliubov quasiparticles then have the conventional quantum numbers.

\begin{figure}
\includegraphics[width=4.0 in]{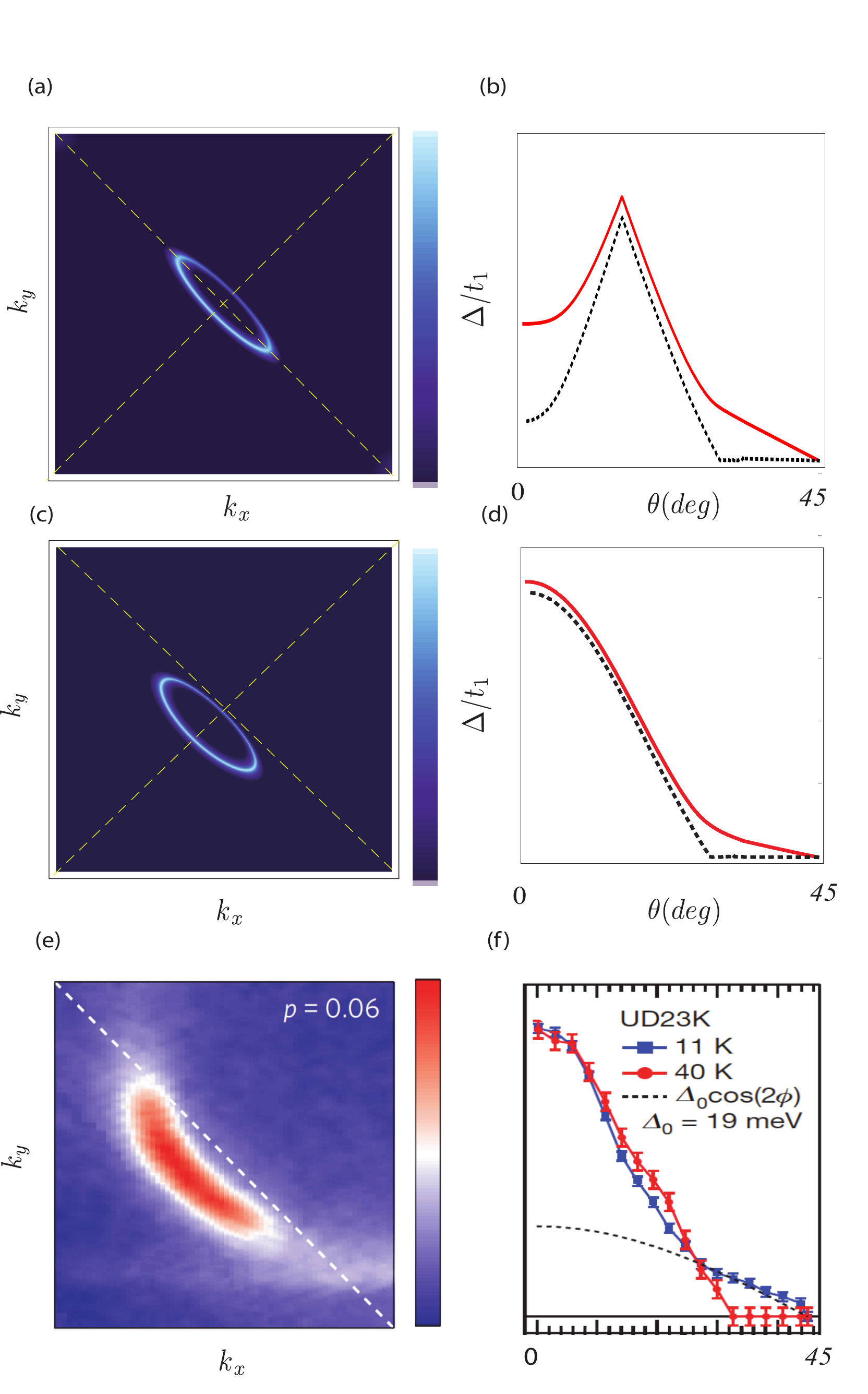}
\caption{Color online: Our new results for the FL* phase (second row), compared with the Hartree-Fock/BCS theory (top row)
and experiments (bottom row). The left panels illustrate Fermi surface structures in the normal state. The right panels
shows the angular dependence of the electron gap in the superconducting states.
(a) Spectral weight of the electron in the normal state with SDW order at wavevector ${\bf K} = (\pi, \pi)$. 
Here we simply apply a potential which oscillates at $(\pi, \pi)$ to the large Fermi surface in the overdoped region.
(b) Minimum electron gap as a function of azimuthal angle in the Brillouin zone. The full (red) line is the result with
 a pairing amplitude $\sim (\cos k_x - \cos k_y )$ co-existing with SDW order, while the dashed (black) line is
 the normal SDW state.
(c) Spectral weight of the electron in the FL* state, with parameters as in Fig.~\ref{Figcase1}; note that the pocket
is no longer centered at $(\pi/2, \pi/2)$. (d) Spectral gap functions in 
the superconducting (full (red) line) and normal (dashed (black) line) states of Fig.~\ref{Figcase1}.
(e) The Fermi pocket from a ARPES experiment \cite{Damascelli}
; related observations appear in Refs.~\onlinecite{tsvelik2,peter3}. 
(f) The dichotomy of the spectral gap function from the
observations of Ref.~\onlinecite{dichot1}. See the text for more details.}
\label{main}
\end{figure}
Our primary results are illustrated in Fig. \ref{main}.
We also show a comparison to a conventional state with co-existing spin density wave (SDW) and 
$d$-wave pairing, and to recent experiments.
The left panels illustrate Fermi surface structures in the normal state. The right panels
show the angular dependence of the electron gap in the superconducting states: for each angle $\theta$,
we determine the minimum electron spectral gap along that direction in the Brillouin zone, and plot
the result as a function of $\theta$.

The results of the traditional Hartree-Fock/BCS theory on SDW order and $d$-wave pairing appear
in (a) and (b). The SDW order has wavevector ${\bf K} = (\pi, \pi)$, and the $d$-wave pairing is the conventional
$(\cos k_x - \cos k_y)$ form. In the normal state, the Fermi pocket is centered at the magnetic Brillouin
zone boundary, as shown in (a).
An important feature of this simple theory is that the state with co-existing SDW and $d$-wave
pairing has
its maximum gap at an intermediate angle, as shown in (b): this reflects the ``hot spots'' which are points on the Fermi surface
linked by the SDW ordering wavevector ${\bf K}$.  No experiment has yet seen such a maximum at an intermediate angle.

One set of our typical results for the FL* theory are shown in 
(c) and (d).
As it was shown in the previous work \cite{pockets}, the normal state in (c) shows a Fermi pocket which is clearly 
not centered the magnetic zone boundary (at $(\pm \pi/2, \pm \pi/2)$); furthermore, its spectral weight is not the same along 
the Fermi surface, and has a arc-like character.
At the same time, (d) shows the angular dependence of the electron gap in the superconducting
state; unlike the SDW theory, this FL* state has a pairing amplitude which is a monotonic
function of angle and has its maximum at the antinodal point. It also shows the   
``dichotomy'' in the gap amplitude between the nodal and anti-nodal regions.
For the purpose of comparison, we illustrate two experimental results in (e) and (f).  
Clearly, our mean-field theory can provide reasonable explanation for the experimental data in both the normal and superconducting
states, and we believe it is 
a candidate for the under-doped cuprate materials.

The structure of this paper is following. 
In Section~\ref{Effective Hamiltonian}, we introduce the normal state Hamiltonian for the fermions, and investigate 
the symmetry transformations of possible pairings. 
We classify possible pairings which preserve full square lattice symmetry, and introduce a low energy effective pairing Hamiltonian.
In Section~\ref{Spectral gaps}, spectral gap functions for various cases are illustrated assuming $d_{x^2-y^2}$ wave pairing. 
It is shown that our model can reproduce the dichotomy behavior, and we compare our theory with the
YRZ model proposed by Yang, Rice and Zhang \cite{YRZ1,YRZ2,YRZ3,YRZ4,YRZ5}, and the related analyses by Wen and Lee \cite{WL1,WL2}.
For completeness, it is shown that $U(1)$ gauge fluctuation can mediate the needed $d$ wave pairing
in Appendix~\ref{Pairing Instability}.

\section{Effective Hamiltonian} \label{Effective Hamiltonian}

The basic setup of the FL* state has been reviewed in some detail in Refs.~\onlinecite{pockets,su2}, and so we will be very
brief here.
The starting point \cite{ss,schulz,wenholon,leeholon,shankar,ioffew,ssss} is to transform 
from the underlying electrons $c_{i \alpha}$ to a rotating reference frame determined
by a matrix $R$ acting on spinless fermions $\psi_p$.
\begin{eqnarray}
c_{i \alpha} = R^i_{\alpha p} \psi_p . \label{cRp}
\end{eqnarray}
$R_{\alpha p}$ is a SU(2) matrix  with $\alpha = \uparrow, \downarrow$ for spin index, $p = \pm$ for gauge index, and we parameterize
\begin{eqnarray}
R^i = \begin{pmatrix} z_{i \uparrow} & -z^*_{i \downarrow} \\  z_{i \downarrow} & z^*_{i \uparrow}   \end{pmatrix}
\end{eqnarray}
with $|z_{i}|^2 =1$ .
In the ACL state, the bosonic $z_\alpha$ and the fermionic $\psi_p$ are assumed to be the independent quasiparticle excitations carrying
spin and charge respectively.
Then we examined the formation of bound states between these excitations. A key result was that was a ``doubling'' of electron-like
quasiparticles, with the availability of two gauge neutral combinations,
\begin{eqnarray}
F_{i \alpha} \sim z_{i \alpha} \psi_{i +} \quad , \quad G_{i \alpha} \sim \varepsilon_{\alpha \beta}z^*_{i \beta} \psi_{i -}.
\end{eqnarray} 
This doubling is a reflection of the `topological order' in the underlying U(1) or $Z_2$ spin liquid; it would
not be present {\em e.g.\/} in a SU(2) spin liquid \cite{su2}.
The $F_{i \alpha}$ and the $G_{i \alpha}$ will be the key actors in our theory of the FL* phase here.
Their effective Hamiltonian is strongly constrained by their non-trivial transformations under the space group of the Hamiltonian, 
which are listed in Table~\ref{table1}.
 \begin{table}[t]
\begin{spacing}{2}
\centering
\begin{tabular}{|c|c|c|c|c|} \hline
 & $T_x$ & $R_{\pi/2}^{\rm dual}$ & $I_x^{\rm dual}$ & $\mathcal{T}$  \\
 \hline  \hline
$ ~F_{\alpha}~$ & $~G_{\alpha}~$ & $ ~G_{\alpha}~$ & $~G_{\alpha}~$  &  $\varepsilon^{\alpha\beta} F_{\beta}^\dagger $ \\
\hline $~G_{\alpha}~$ & $~F_{\alpha}~$ & $~F_{\alpha}~$ & $~F_{\alpha}~$  & $
\varepsilon^{\alpha\beta} G_{\beta}^\dagger$ \\ \hline \hline
$ ~C_{\alpha}~$ & $~C_{\alpha}~$ & $ ~C_{\alpha}~$ & $~C_{\alpha}~$  &  $\varepsilon^{\alpha\beta} C_{\beta}^\dagger $ \\
\hline $~D_{\alpha}~$ & $~D_{\alpha}~$ & $~D_{\alpha}~$ & $~D_{\alpha}~$  & $
\varepsilon^{\alpha\beta} D_{\beta}^\dagger$ \\ \hline
\end{tabular}
\end{spacing}
\caption{Transformations of the lattice fields under square
lattice symmetry operations.  $T_x$: translation by one lattice
spacing along the $x$ direction; $R_{\pi/2}^{\rm dual}$: 90$^\circ$
rotation about a dual lattice site on the plaquette center
($x\rightarrow y,y\rightarrow-x$); $I_x^{\rm dual}$: reflection
about the dual lattice $y$ axis ($x\rightarrow -x,y\rightarrow y$);
$\mathcal{T}$: time-reversal, defined as a symmetry (similar to
parity) of the imaginary time path integral. Note that such a
$\mathcal{T}$ operation is not anti-linear. \cite{pockets} }
\label{table1}
\end{table}

From these symmetry transformations, we can write down the following effective Hamiltonian \cite{pockets}
\begin{eqnarray}
H_{tot} &=& H_{0} + H_{int} \nonumber \\
H_{0} &=& -\sum_{ij} t_{ij} (F_{i \alpha}^{\dagger} F_{j \alpha}+ G_{i \alpha}^{\dagger} G_{j \alpha}) + \lambda \sum_{i} (-1)^{i_x+i_y}(F_{i \alpha}^{\dagger} F_{i \alpha}- G_{i \alpha}^{\dagger} G_{i \alpha})  \nonumber \\
&&   - \sum_{i<j} \tilde t_{ij} ~ (F_{i \alpha}^{\dagger} G_{j \alpha}+ G_{i \alpha}^{\dagger} F_{j \alpha}).
\label{hamiltonian}
\end{eqnarray}
Here $t_{ij}$ is taken to be similar to the bare electron dispersion,
characterizing the Fermi surface in the over-doped region; $\lambda$ is a potential due to the local antiferromagnetic order;
and $\tilde t_{ij}$ is the analog of the Shraimain-Siggia term \cite{ss} which couples the two species of electron-like
quasiparticles $F$ and $G$ to each other; it is this term which is responsible for shifting the center of the pocket Fermi surfaces
in the normal state away from the magnetic Brillouin zone boundary. $H_{int}$ is the invariant interaction Hamiltonian:
there could be many interaction channels, which induces superconductivity of the $(F,G)$ particles, such as negative contact interaction, interaction with other order parameters, and the gauge field fluctuation. 
In this paper, we do not specify particular interaction and we assume that pairings are induced.
Then we focus on studying properties of possible pairings and their consequences on physical quantities such as spectral gaps. 
In Appendix~\ref{Pairing Instability}, we illustrate one possible channel to achieve such superconductivity. 

For some of our computations, it is more convenient to use an alternative basis for the fermion operators
\begin{eqnarray}
C_{i,\alpha} &=& \frac{1}{\sqrt{2}} (F_{i,\alpha} + G_{i,\alpha}) \quad , \quad D_{i,\alpha} = (-1)^{i_x+i_y} \frac{1}{\sqrt{2}} (F_{i,\alpha} - G_{i,\alpha}).
\end{eqnarray}
The $C$ and $D$ fermions have the same space-group transformation properties as the physical electrons.
Then, the Hamiltonian becoms 
\begin{eqnarray}
H_{0}&=& \sum_{{\bf k},\alpha} \begin{pmatrix} C_{{\bf k},\alpha} \\  D_{{\bf k},\alpha}  \end{pmatrix}^{\dagger} \begin{pmatrix} \epsilon_c ({\bf k}) &\lambda \\  \lambda & \epsilon_d ({\bf k}) \end{pmatrix} \begin{pmatrix} C_{{\bf k},\alpha} \\  D_{{\bf k},\alpha} \end{pmatrix} 
\end{eqnarray} 
We chose  energy dispersion's forms following the previous work\cite{pockets}, with $\epsilon ({\bf k})$ a Fourier transform
of $t_{ij}$ and $\tilde \epsilon ({\bf k})$ a Fourier transform of $\tilde t_{ij}$, and ${\bf K} = (\pi, \pi)$:
\begin{eqnarray}
\epsilon({\bf k}) &=& -2t_1 (\cos k_x +\cos k_y) + 8 t_2 \cos k_x \cos k_y -2 t_3 (\cos 2k_x +\cos 2k_y)  \nonumber \\
\tilde \epsilon({\bf k}) &=& -\tilde t_0 -2\tilde t_1 (\cos k_x +\cos k_y) + 8 \tilde t_2 \cos k_x \cos k_y -2 \tilde t_3 (\cos 2k_x +\cos 2k_y) \nonumber \\
\epsilon_c({\bf k}) &=& \epsilon({\bf k})+\tilde \epsilon({\bf k}) - \mu \nonumber \\ 
\epsilon_d({\bf k}) &=&  \epsilon({\bf k}+ {\bf K})-\tilde \epsilon({\bf k}+{\bf K}) - \mu . 
\end{eqnarray}

The $C$ and $D$ particles have spin and electric charges like electrons.
Therefore, any linear combination can be a candidate for the physical electron degree of freedom. 
In the previous work \cite{pockets}, we matched the $C$ particles to the electrons of large Fermi 
surface state without antiferromagnetism; following this, for simplicity we will take the $C$ to be the physical electron,
but our results do not change substantially with other linear combinations.
Then the $D$ particles are emergent fermion induced by fluctuating SDW order. 
Note that the $C,D$ particles live in the {\em full\/} first Brillouin zone of the square lattice,
and not the magnetic Brillouin zone.

Issues related to the Luttinger theorem were discussed in previous work \cite{rkk1,rkk2,pockets}.
The total area of the Fermi pockets described by $H_0$ is precisely $x$, the dopant hole density.
Here the area is to be computed over the full first Brillouin zone of the square lattice,
as the full square lattice symmetry is preserved by our model. Also note that our phenomenological 
Hamiltonian $H_0$ has been designed to apply only to low energy excitations near the Fermi surface.
However, rather than focusing on these momentum space regions alone, considerations of symmetry
are far simpler if we define the dispersion in real space on the underlying square lattice, as we have done here.
For this somewhat artificial lattice model, as discussed in Ref.~\onlinecite{pockets},
the total fermion density on each site $i$ is
\begin{equation}
\sum_\alpha \left \langle C_{i,\alpha}^\dagger C_{i , \alpha} + D_{i,\alpha}^\dagger D_{i , \alpha} \right\rangle = 
\sum_\alpha \left \langle F_{i,\alpha}^\dagger F_{i , \alpha} + G_{i,\alpha}^\dagger G_{i , \alpha} \right\rangle =2-x
\end{equation}
The traditional Luttinger theorem measures electron number modulo 2, and so it should now
be clear that occupying the independent electron states of the lattice $H_0$ will yield
a Fermi surface with the desired area of $x$.

Before proceeding further, let us review the above discussion. 
We started our theory with electrons in one band, and considered spin density wave fluctuation. 
The strong fluctuation induced particle fractionalization, and bound states whose degree of freedoms are doubled appeared. 
The resulting phase is nothing but the FL* we introduced above.
Therefore, the ACL phase provides a natural way to connect the FL* with one band theory.

 To study superconductivity of the FL* phase, let us consider invariant pairing operators.
 With the $(F,G)$ particles, there are many possible combinations in principle. 
 However, it is more convenient to work in terms of the $C$ and $D$ particles because they transform
 just like electrons under the symmetry operation. So we can write down the 4 pairing operators
\begin{eqnarray}
O^{c}_{\Delta}({i,j}) & =&\varepsilon^{\alpha\beta} C_{i,\alpha}C_{j,\beta} ,\quad O^{d}_{\Delta}({i,j}) =\varepsilon^{\alpha\beta} D_{i,\alpha}D_{j,\beta}  \nonumber \\
O^{cd}_{\Delta}({i,j})& =&\varepsilon^{\alpha\beta} C_{i,\alpha}D_{j,\beta} ,\quad O^{dc}_{\Delta}({i,j}) = \varepsilon^{\alpha\beta} D_{i,\alpha}C_{j,\beta} 
  \label{cdtoAB} 
\end{eqnarray} 
Note that we only consider even parity pairing, and there are only 
three pairings, $O^c, O^d, O^{cd}+O^{dc}$ (see Appendix~\ref{app:pairing}).

\section{Spectral Gap} \label{Spectral gaps}

Throughout this paper, we assume that all pairings are $d$ wave, more specifically, $d_{x^2-y^2}$. 
The assumption of the $d$ wave pairings can be realized by the gauge fluctuation (see Appendix~\ref{Pairing Instability}) 
or by other channels like conventional spin density wave fluctuations. 
Then, with the pairing amplitudes as in Eq.~(\ref{cdtoAB}), we can write down the mean field Hamiltonian  
\begin{eqnarray} 
 H_{tot}^{MF}  &=& H_{0} + H_{\Delta}^{MF} \nonumber \\
 &=& \sum_{{\bf k}} 
\begin{pmatrix} C_{{\bf k},\uparrow}^{\dagger} \\ C_{ -{\bf k},\downarrow} \\D_{{\bf k},\uparrow}^{\dagger} \\  D_{-{\bf k},\downarrow}  \end{pmatrix} 
\begin{pmatrix} \epsilon_c ({\bf k}) & -\Delta_c({\bf k}) &\lambda &  -\Delta_X({\bf k}) \\
 -\Delta_c({\bf k})^{*} &  -\epsilon_c ({\bf k})  & - \Delta_X({\bf k})^{*}& - \lambda  \\ 	
\lambda &  -\Delta_X({\bf k}) &  \epsilon_d ({\bf k})  & - \Delta_d({\bf k}) \\ 
 -\Delta_X({\bf k})^{*}& - \lambda  &- \Delta_d({\bf k})^{*}&  -\epsilon_d ({\bf k}) 
\end{pmatrix} 
\begin{pmatrix} C_{{\bf k},\uparrow} \\ C_{ -{\bf k},\downarrow}^{\dagger} \\D_{{\bf k},\uparrow}\\  D_{-{\bf k},\downarrow}^{\dagger}   \end{pmatrix} \label{MF}
\end{eqnarray} 
where $\Delta_c$ is the Fourier transform of $O^{c}_{\Delta}$, $\Delta_d$ is the Fourier transform of $O^{d}_{\Delta}$,
and $\Delta_X$ is the Fourier transform of $O^{cd}_{\Delta} + O^{dc}_{\Delta}$. For their wavevector dependence
we take the forms
\begin{equation}
\frac{\Delta_c ({\bf k})}{\Delta_{c0}} = \frac{\Delta_d ({\bf k})}{\Delta_{d0}} = \frac{\Delta_X ({\bf k})}{\Delta_{X0}}
 = \cos k_x - \cos k_y
\end{equation}
where $\Delta_{c0}$, $\Delta_{d0}$ and $\Delta_{X0}$ are the respective gap amplitudes.

In principle, we could determine these pairing amplitudes from solving a set of BCS-like self-consistency 
equations. However, in the absence of detailed knowledge of the pairing interactions, we will
just treat the $\Delta_{c0}$, $\Delta_{d0}$ and $\Delta_{X0}$ as free parameters.
In other words, we are in the deep superconducting phase with adjusted parameters. 
Then our task is to study spectral gap behaviors with given band structures and pairings.
More technically, the Green's function of the $C$ particle, which determines the electron properties, are studied focusing on the pole of  the $C$ particles' Green's function. 
The pole basically contains information about the electron's dispersion relation, and its minimum determines spectral gap properties. The latter is defined as the minimum gap along a line from the Brillouin zone center at an angle $\theta$:
thus the nodal point is at $\theta = \pi/4$, and the anti-nodal point at $\theta = 0$.

Although we have three free gap parameters, our results are quite insensitive to their values. For simplicity
we will mainly work (in Sections~\ref{sec:singleI} and~\ref{sec:singleII})
with the case with a single gap parameter $\Delta_{c0} \neq 0$, and others are set to zero
$\Delta_{d0} = \Delta_{X0} = 0$. We will briefly consider the case with multiple gap parameters in Section~\ref{sec:multiple},
and find no significant changes from single gap case.

\subsection{Single Gap : case I}
\label{sec:singleI}

\begin{figure}
\includegraphics[width=4.0 in]{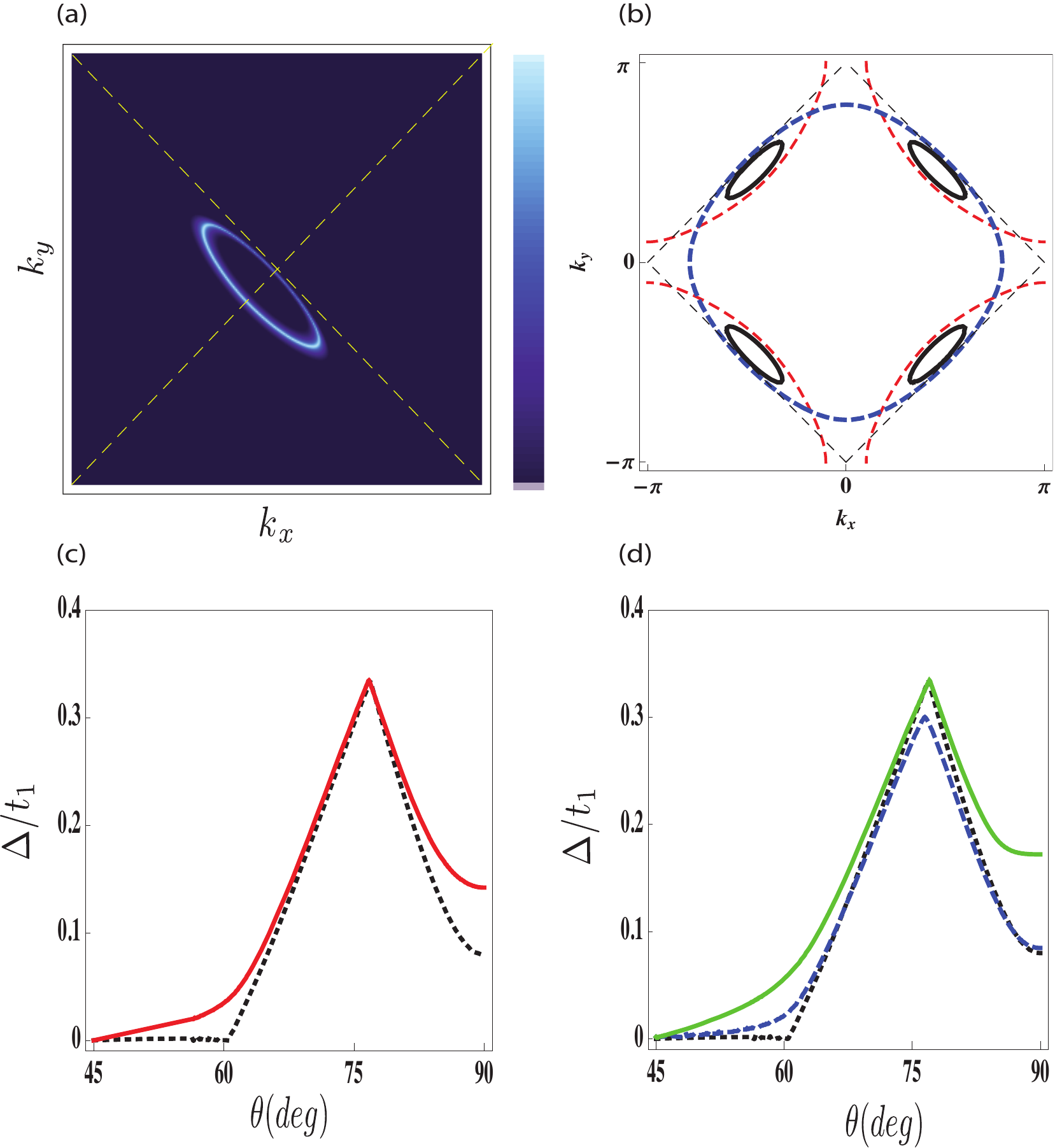}
\caption{Color online: Spectral gap functions and the Fermi surfaces with  the Case I, ($ t_2 = 0.15 t_1$,  $t_3 = -0.3 t_2$, $\tilde{t}_1 = -0.25 t_1$, 
$\tilde{t}_2 = 0$,  $\tilde{t}_3 = 0$, $\tilde{t}_0 = -0.3 t_1$, $\mu = -0.6 t_1$, $\lambda = 0.4 t_1$. 
(a) The spectral weight of the electron Green function with the relaxation time $\tau t_1 =200$. 
(b) Fermi surfaces of $\epsilon_c ({\bf k}) $ (dashed inner (red)) , $\epsilon_d ({\bf k})$ (dashed outer (blue)), 
and the eigenmodes (thick (black)) of $H_0$.  The dotted line is the magnetic zone boundary. 
(c) The spectral gap function with and without $\Delta_c$. The dotted (black) line is for the normal case. 
The thick (red) line is for superconducting state with $\Delta_{c0}=0.1 t_1$. 
(d) The spectral gap function with and without $\Delta_{d,X}$. The dotted (black) line is for the normal case. The thick (green) line is for the superconducting state with $\Delta_{X0}=0.1 t_1$.
The dashed (blue) line is the superconducting state with $\Delta_{d0}=0.1 t_1$.  }
\label{Figcase2}
\end{figure}

We consider the case with $ t_2 = 0.15 t_1$,  $t_3 = -0.3 t_2$, $\tilde{t}_1 = -0.25 t_1$, 
$\tilde{t}_2 = 0$,  $\tilde{t}_3 = 0$, $\tilde{t}_0 = -0.3 t_1$, $\mu = -0.6 t_1$, $\lambda = 0.4 t_1$ in Fig.~\ref{Figcase2}. 
In (a), the calculated spectral weight of the C particle is illustrated following the previous paper.\cite{pockets}
The shape is obviously pocket-like, but its spectral weight depends on position on the Fermi surface. 
In (b), we illustrate the bare energy Fermi surfaces and their eigenmode Fermi surface. 
Note that the two bare energy bands ($\epsilon_{c,d} ({\bf k})$) are different from the usual SDW formations with Brillouin zone folding. 
In the latter, there is only one electron band, and SDW onset divides the Brillouin zone two pieces ($\epsilon( {\bf k}),
\epsilon( {\bf k}+{\bf K})$).
But in our case, the two bands have different energy spectrums of the electron-like particle ($C$) and the emergent particle ($D$). 
And $\lambda$ determines mixing energy scale between the $C$ and $D$ particles.
 
In (c), the spectral gap function with and without a given pairing, $\Delta_c$ is illustrated. 
Near the node, it is obvious that the pairing gap contributes to the spectral gap in a $d$ wave pairing way as expected. However, between the node and anti-node, there is a huge peak. 
The peak position is nothing but the mixing point between $C,D$ particles. 
Therefore, the peak exists whether there is a pairing or not. 
Near the anti-node, the spectral gap is bigger than the near-node's but much smaller than the mixing point peak. 
It indicates there is tendency to make electron pockets near the anti-node. 
For example,  if we decrease the magnitude of $\lambda$, which basically represent the mixing energy scale, then the gap near the anti-node becomes smaller, and eventually  the electron-like pockets appears near the anti-node  with the pre-existing hole type pockets. 
(See the Appendix)
Note that this situation is formally the same as the pairing with the SDW fluctuation mediating pairing case (see Fig. \ref{main}).
The ``hot spot'' between the node and the anti-node has the largest gap magnitude, which corresponds to our mixing point.
Such a spectral gap behavior is not the experimentally observed one. 
Therefore, we cannot have the needed dichotomy near the anti-node in this case; the anti-nodal gap is always smaller than the one of the maximum mixing point.
Following the similar reasoning, the experimentally observed dichotomy does not appear in the conventional SDW theory unless additional consideration beyond mean-field theory is included.   
In (d), we illustrate other pairing cases ($\Delta_{d,X}$). 
As we can see, the role of the pairings are similar to the conventional one ($\Delta_c$), and qualitatively they are the same. 
Therefore, it is not possible to achieve the observed dichotomy by considering the exotic pairings. 
They cannot push the maximum peak of the normal state to the anti-nodal region. 

The message of this calculation is simple. 
With the band structure we considered here, the observed dichotomy in the spectral gap function cannot be obtained, 
even though the normal state can explain experimentally observed Fermi surface structures. 
Moreover, it also implies that it is difficult to explain the observed dichotomy with the Hartree-Fock/BCS mean-field theory
of the Fermi liquid. 

However, we now show how our FL* theory gets a route to explain the dichotomy below.

\subsection{Single Gap : case II}
\label{sec:singleII}

\begin{figure}
\includegraphics[width=4.0 in]{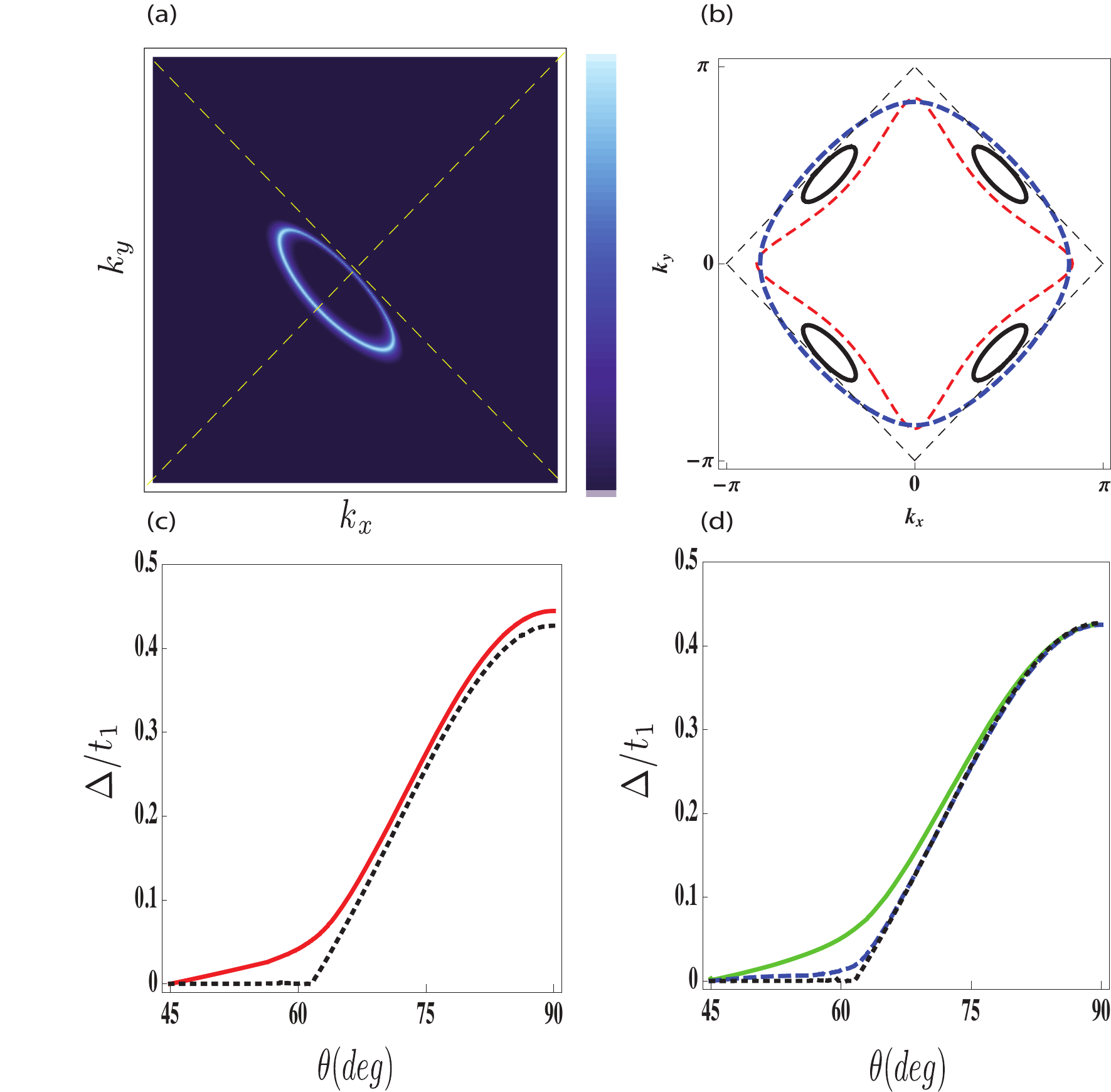}
\caption{Color online: Spectral gap functions and the Fermi surfaces for 
the Case II ($ t_2 = 0.15 t_1$,  $t_3 = -0.3 t_2$, $\tilde{t}_1 = -0.25 t_1$, 
$\tilde{t}_2 = 0$,  $\tilde{t}_3 = 0$, $\tilde{t}_0 = -0.3 t_1$, $\mu = -0.8 t_1$, $\lambda = 0.6 t_1$). 
Note that the only change from Fig.~\ref{Figcase2} is in the values of $\mu$ and $\lambda$.
(a) The spectral weight of the electron Green function with the relaxation time $\tau t_1 =200$. 
(b) Fermi surfaces of $\varepsilon_c$ (dashed inner (red)) , $\varepsilon_d$ (dashed outer (blue)), and the eigenmodes
(thick (black)) of $H_0$.  The dotted line is the magnetic zone boundary. 
(c) The spectral gap function with and without $\Delta_c$. The dotted (black) line is for the normal case
with $\Delta_c = 0$. The thick (red) line is the superconducting state with $\Delta_{c0}=0.1 t_1$. 
(d) The spectral gap function with and without $\Delta_{d,X}$. The dotted (black) line is for the normal case. The thick
(green) line has $\Delta_{X0}=0.1 t_1$. The dashed (blue) line has $\Delta_{d0}=0.1 t_1$.  }
\label{Figcase1}
\end{figure}

In Fig. \ref{Figcase1} we illustrate the case with $ t_2 = 0.15 t_1$,  $t_3 = -0.3 t_2$, $\tilde{t}_1 = -0.25 t_1$, 
$\tilde{t}_2 = 0$,  $\tilde{t}_3 = 0$, $\tilde{t}_0 = -0.3 t_1$, $\mu = -0.8 t_1$ , $\lambda = 0.6 t_1$.
These parameters are as in Section~\ref{sec:singleI}, except that the values of $\mu $ and $\lambda$ have changed.
As we discuss below, this changes the structure of the dispersion of the `bare' $C$ and $D$ particles in a manner which
leaves the normal state Fermi surface invariant, but dramatically modifies the spectral gap in the superconducting state.

As we can see in (a), the calculated spectral weight of the C particle is qualitatively the same as the Fig. \ref{Figcase2}'s. The shape is obviously pocket-like, and its spectral weight also depends on position of the Fermi surface similarly.
Therefore, in the normal state, there is no way to distinguish the two cases because the low energy theory are all determined by the Fermi pocket structures. 
However, in (b), the bare energy Fermi surfaces of $\epsilon_c ({\bf k})$ and $\epsilon_d ({\bf k})$  
are clearly different from the previous one's.
Even though the bare Fermi surfaces look unfamiliar, they are irrelevant for the observed Fermi surface which is determined by the eigenmodes of $H_0$ (black line), and which is qualitatively the same as the Case I. 

We illustrate our spectral gap behavior with and without the pairing, $\Delta_c$, in (c), which was already shown in the introductory section. 
Without the given pairing, the normal state has the finite gapless region where the pockets exist, and there is a stable spectral gap in the anti-node. 
It is easy to check the anti-nodal gap depends on the mixing term, $\lambda$, between the $C$ and $D$ particles. 
With the pairing, the Fermi pockets are gapped and only the node remains gapless. 
The spectral gap function has expected $d$ wave type gap near the node, and the observed dichotomy  is clearly shown.
Therefore, the origin of the two gaps are manifest; the nodal gap is obviously from the $C$ particle pairing and the anti-nodal gap is originated from the mixing term, which is inherited from the spin-fermion interaction term. 
In (d), we illustrate other exotic pairings ($\Delta_{d,X}$). 
As we can see, role of the pairings are similar to the conventional pairing ($\Delta_c$), and qualitatively they are the same. 
So, there is no way to distinguish what pairings are dominant only by studying spectral gaps. 

Now let us compare our results to the ones of the YRZ model\cite{YRZ1,YRZ2,YRZ3,YRZ4,YRZ5}.
In the YRZ model, based on a specific spin liquid model, the pseudo-gap behavior is pre-assumed by putting an 
explicit $d_{x^2-y^2}$ gap function in the spectrum, which means the characteristic of the anti-nodal gap is another input parameter. 
With the two $d$ wave gaps (pairing and pseudo-gap), the experimental results were fitted. 

In our FL* theory, the anti-nodal gap behavior is determined by the interplay between $\lambda$ and the bare spectrum
$\epsilon_{c,d} ({\bf k})$
Indeed, the pseudo-gap corresponding term, $\lambda$, is $s$ wave type in terms of YRZ terminology. 
The $\lambda$ term represents local antiferromagnetism, and 
this `competing' order which plays a significant role in the anti-nodal gap.
The parameter $\lambda$ is just input for making the Fermi pockets in the normal state with other dispersion parameters.
As mentioned before, it explains the distinct origins of the nodal and anti-nodal gaps.  
Also, although our theory contains other pairings, $\Delta_{d,X}$, we did not need that freedom to obtain consistency
with experimental observations.

Of course, non-local terms of $\lambda$ could be considered.
And it is easy to show that  the $d_{x^2-y^2}$ like terms are not allowed because of the rotational symmetry breaking. 
Putting the non-local $\lambda$ term is secondary effect, and we do not consider it here.

\subsection{Multiple gaps}
\label{sec:multiple}

\begin{figure}
\includegraphics[width=3.0 in]{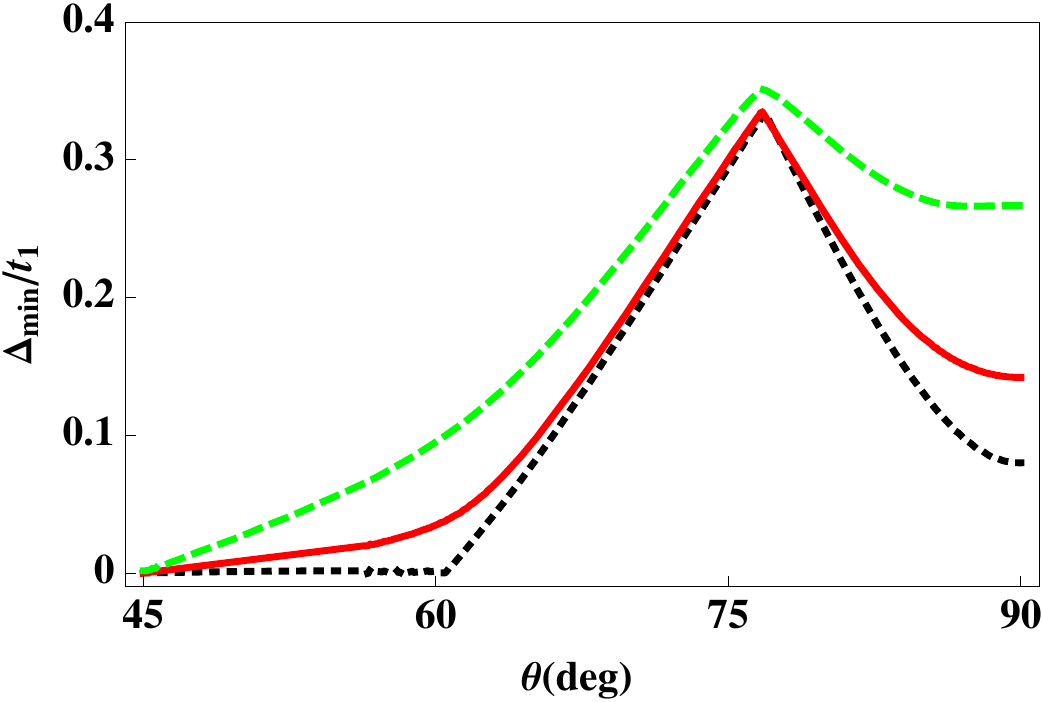}
\includegraphics[width=3.0 in]{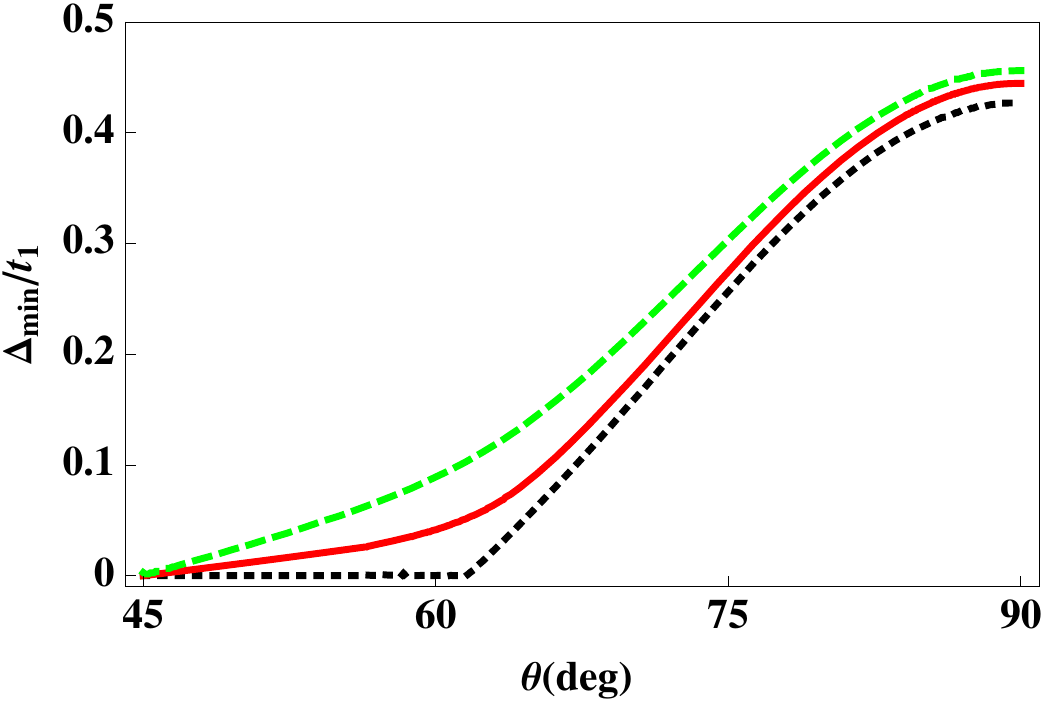}
\caption{Multiple gaps. 
The left panel is the same as the Fig. \ref{Figcase2} with two superconducting gaps $\Delta_{d0} = 0.3 t_1$ 
and $\Delta_{c0} =0.1 t_1$. 
And the right panel is the same as the Fig. \ref{Figcase1}  with  $\Delta_{d0} = 0.3 t_1$ and $\Delta_{c0} = 0.1 t_1$. 
In both, the dashed (green) line is with the two gaps. And the plain and dotted lines are the same as the previous plots.}
\label{multiple1}
\end{figure}

So far, we have only considered the cases with one pairing gap.
Of course, multiple gaps are possible and we illustrate possible two cases in Fig.~\ref{multiple1}, which contain 
$\Delta_{c,d}$ with the two normal band structures. 
 Here, we choose the same phase in both pairings. 
The spectral gap behaviors are not self-destructive, which means the magnitude of spectral gap with two pairings is bigger than the one with the single pairings. 
One comment is that even multiple gaps do not change qualitative behavior of the spectral gap functions, which means that the Case I could not have the observed dichotomy even with the multiple gaps. 

\begin{figure}
\includegraphics[width=3.0 in]{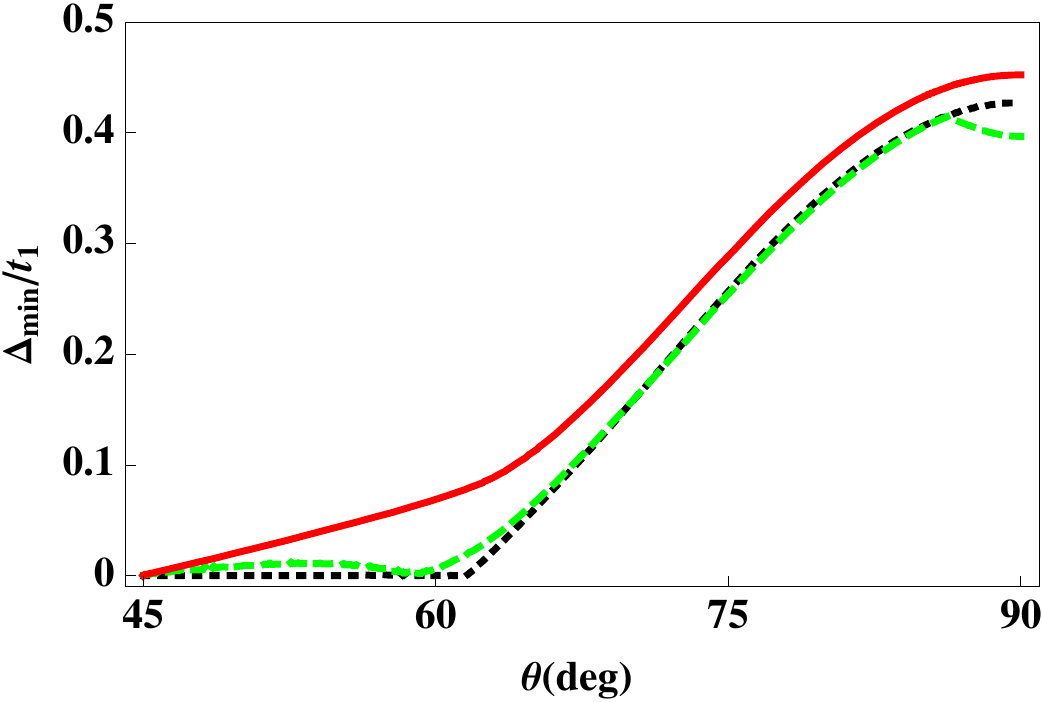}
\caption{Multiple gaps with the relative phase difference. 
Details are the same as Fig.~\ref{Figcase1}.
The red line is for two superconducting gaps with the same sign gaps, $\Delta_{d0} =0.1 t_1$ and $\Delta_{c0}  =0.1 t_1$. 
And the green line is for the opposite sign gaps, $\Delta_{d0} =-0.1 t_1$ and $\Delta_{c0} =0.1 t_1$.  }
\label{multiple2}
\end{figure}

In Fig. \ref{multiple2}, two pairings with the opposite sign are illustrated. 
Clearly, we can see the self-destructive pattern with the same gap magnitudes.
Even a node appears beyond the nodal point. 
Therefore, it is clear that the relative phase between two pairings plays an important role to determine the gap spectrum. 

\section{Conclusions} \label{Conclusion}

This paper has presented a simple phenomenological model for pairing in the underdoped cuprates,
starting from the FL* normal state described in Ref.~\onlinecite{pockets}. This is an exotic normal
state in which the Cu spins are assumed to form a spin liquid, and the dopants then occupy states
with electron-like quantum numbers. A key feature of this procedure \cite{rkk1}, is that there is a
`doubling' of the electron-like species \cite{rkk1} 
available for the dopants to occupy: this appears to be a generic
property of such doped FL* states. 

Our previous work \cite{pockets} showed how this model could easily capture the Fermi surface structure of the
underdoped normal state. In particular, a mixing between the doubled fermion $F$ and $G$ species 
from the analog of the `Shraiman-Siggia' term \cite{ss}
led to Fermi pockets which were centered away from the antiferromagnetic Brillouin zone boundary.

Here we considered the paired electron theory, assuming a generic $d$-wave
gap pairing of the $\cos k_x - \cos k_y$ variety. Despite this simple gap structure, we found two distinct types
of electron spectral gaps in this case, illustrated in Figs.~\ref{Figcase2} and~\ref{Figcase1}. The distinction arose
mainly from the strength of a parameter, $\lambda$, determining the strength of the local antiferromagnetic order.

For weaker local antiferromagnetic order, and with a normal state Fermi surface as in Fig.~\ref{Figcase2}a, the angular
dependence of the gap had a strong maximum near the intermediate ``hot spot'' on the underlying Fermi surface.
A similar structure is seen in the traditional Hartree-Fock/BCS theory of SDW and $d$-wave pairing on a normal Fermi liquid,
and this structure is incompatible with existing experiments.

For stronger local antiferromagnetic order, we were able to maintain the normal state Fermi surface as in Fig.~\ref{Figcase1}a,
but then found a gap function which had the form shown in Figs.~\ref{Figcase1}c,d, which displays the `dichotomy'
of recent observations. 
Thus in this theory, it is the fluctuating local antiferromagnetism which controls the dichotomy.
  
Finally, we compare our theory with model proposed by Yang, Rice, and Zhang \cite{YRZ1,YRZ2,YRZ3,YRZ4,YRZ5},
and closely related results of Wen and Lee \cite{WL1,WL2}.
Their phenomenological form of the normal state electron Green's function has
qualitative similarities to ours \cite{pockets}, but there are key differences in detail:\\ 
({\em i\/}) The `back end'
of the YRZ hole pocket is constrained to be at $(\pi/2,\pi/2)$, while there is no analogous pinning
in our case.\\ 
({\em ii\/}) The electron spectral weight vanishes in the YRZ theory at $(\pi/2, \pi/2)$, while our theory
has a small, but non-zero, spectral weight at the back end.\\
({\em iii\/}) Our theory allows for a state with both electron and hole and pockets, while only hole pockets
are present in the YRZ theory.\\
These differences can be traced to the distinct origins of the `pseudogap' in the two theories. Our pseudogap 
has connections to local
antiferromagnetism which fluctuates in orientation while suppressing topological defects. Pairing correlations 
also play an important role in the pseudogap, but these are neglected in our present mean-field
description: these were examined in our previous fluctuation analyses of the ACL \cite{gs,superACL}. 
The YRZ pseudogap is due to a $d$-wave 
`spinon pairing gap' in a resonating valence bond spin liquid. All approaches have a similar
transition to superconductivity, with a $d$-wave pairing gap appearing over the normal state spectrum, and a nodal-anti-nodal
dichotomy: thus any
differences in the superconducting state can be traced to those in the normal state.

The differences between our normal state theory with bosonic spinons, and other work based upon fermionic spinons \cite{YRZ1,YRZ2,YRZ3,YRZ4,YRZ5,WL1,WL2,RW1,RW2,RW3} become more pronounced
when we consider a transition from the normal state to a state with long-range antiferromagnetic order.
In our theory, such a transition is naturally realized by condensation of bosonic spinons, with universal characteristics
discussed earlier \cite{senthil,css}. Such a natural connection to the antiferromagnetically ordered state is not
present in the YRZ theory.

\acknowledgements

We thank P.~Johnson, M.~Randeria, T.~M.~Rice, and T.~Senthil for useful discussions.
This research was supported by the National Science Foundation under grant DMR-0757145 and by a MURI grant from AFOSR.
 
\appendix
 
\section{Electron pockets}
\begin{figure}
\includegraphics[width=4.0 in]{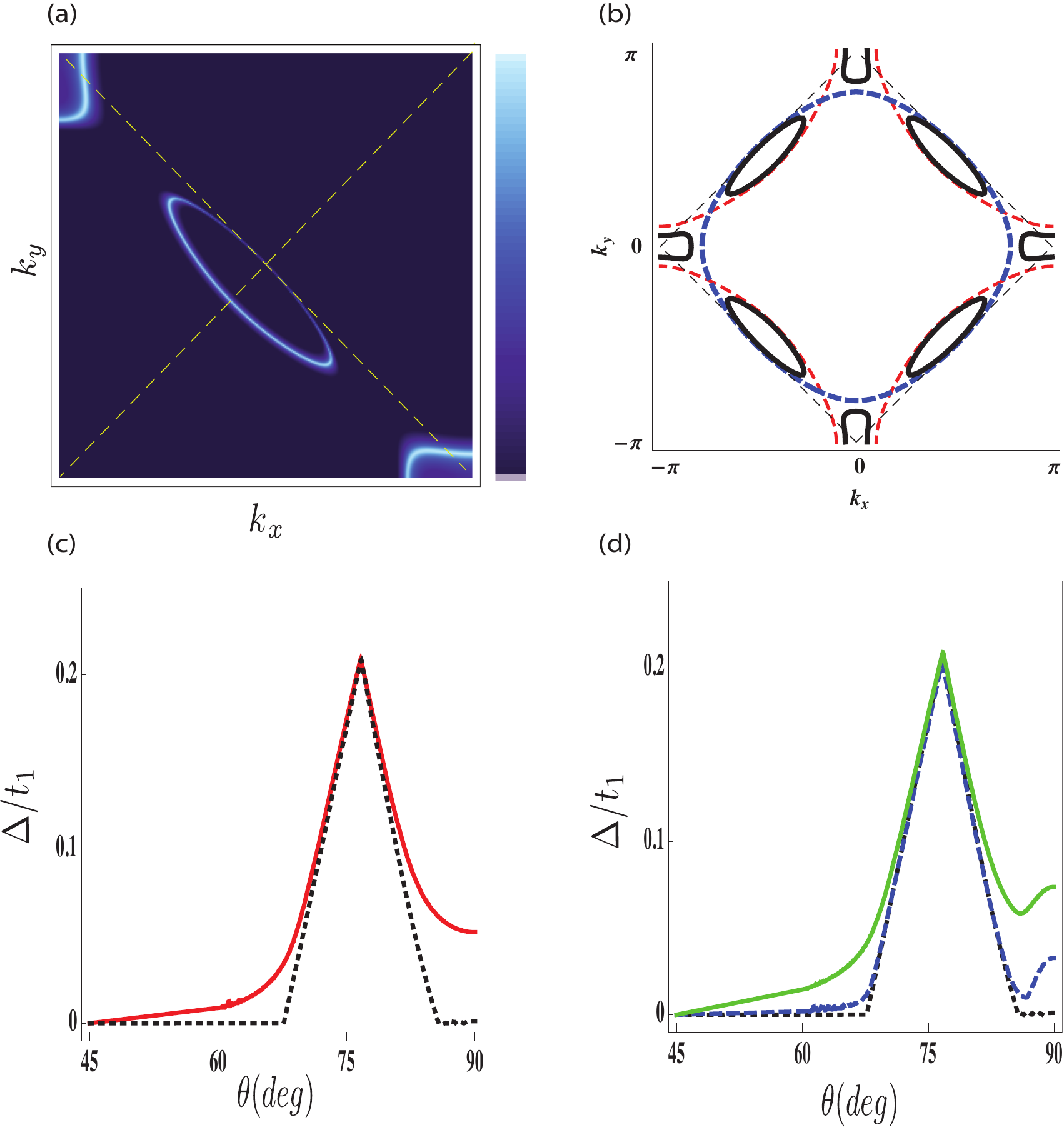}
\caption{Color online: Spectral gap functions and the Fermi surfaces for 
the Case II ($ t_2 = 0.15 t_1$,  $t_3 = -0.3 t_2$, $\tilde{t}_1 = -0.25 t_1$, 
$\tilde{t}_2 = 0$,  $\tilde{t}_3 = 0$, $\tilde{t}_0 = -0.3 t_1$, $\mu = -0.6 t_1$, $\lambda = 0.25 t_1$). 
Note that the only change from Fig.~\ref{Figcase1} is in the value of $\lambda$.
(a) The spectral weight of the electron Green function with the relaxation time $\tau t_1 =200$. 
(b) Fermi surfaces of $\varepsilon_c$ (dashed inner (red)) , $\varepsilon_d$ (dashed outer (blue)), and the eigenmodes
(thick (black)) of $H_0$.  The dotted line is the magnetic zone boundary. 
(c) The spectral gap function with and without $\Delta_c$. The dotted (black) line is for the normal case
with $\Delta_c = 0$. The thick (red) line is the superconducting state with $\Delta_{c0}=0.05 t_1$. 
(d) The spectral gap function with and without $\Delta_{d,X}$. The dotted (black) line is for the normal case. The thick
(green) line has $\Delta_{X0}=0.05 t_1$. The dashed (blue) line has $\Delta_{d0}=0.05 t_1$.  }
\label{FigEpockets}
\end{figure}

We consider the case with $ t_2 = 0.15 t_1$,  $t_3 = -0.3 t_2$, $\tilde{t}_1 = -0.25 t_1$, 
$\tilde{t}_2 = 0$,  $\tilde{t}_3 = 0$, $\tilde{t}_0 = -0.3 t_1$, $\mu = -0.6 t_1$, $\lambda = 0.25 t_1$ in Fig. \ref{FigEpockets}.
These parameters are as in Section~\ref{sec:singleI}, except that the value of $\lambda$ has lowered.
In other words, the `bare' spectrums are the same, but electron pockets near the anti-node appear due to the low mixing term.

\begin{figure}
\includegraphics[width=4.0 in]{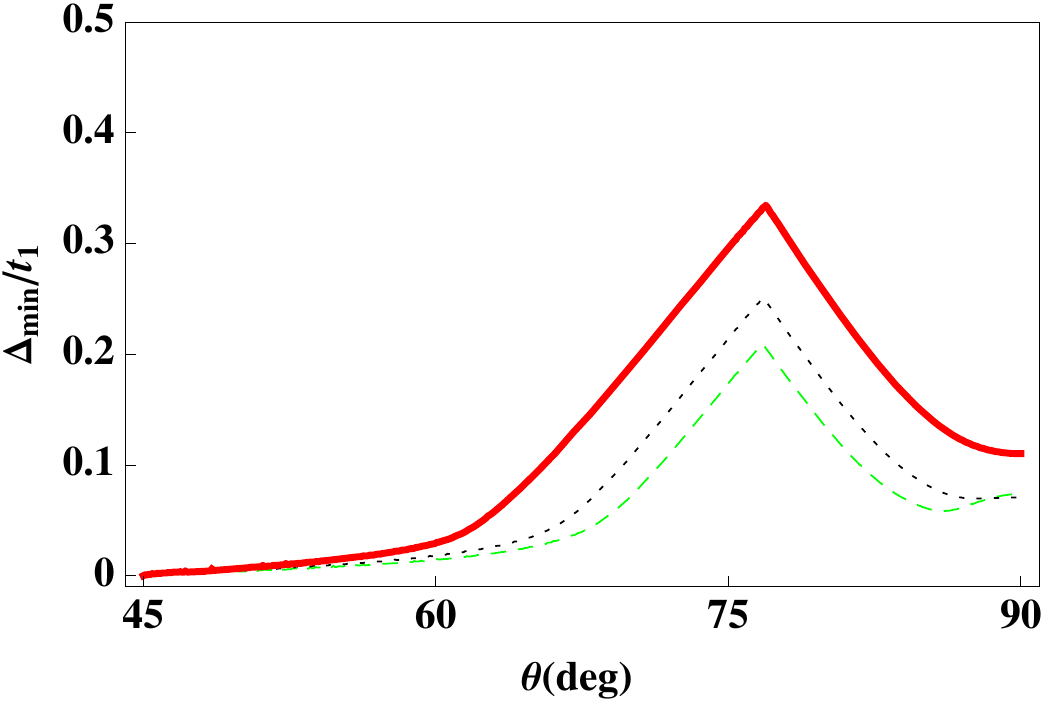}
\caption{Color online : Spectral gap behaviors varying with $\lambda$. The thick(red), dotted(black) and dashed(green) lines are for $\lambda / t_1=0.4, 0.3, 0.25$ with the same pairing magnitude, $\Delta_X =0.05 t_1$. } \label{xseries}
\end{figure}

As we can see in (a), the calculated spectral weight of the C particle shows the hole and electron pockets with different spectral weights. 
We illustrate our spectral gap behavior with and without the pairing in (c) and (d).
Without pairings, the normal state has the finite gapless region where the pockets exist, and there is an intermediate region peak similar to the SDW case. 
With pairings, the Fermi pockets are gapped and only the node remains gapless. 
The spectral gap function shows similar behavior as in our case I. 
In Fig. \ref{xseries}, spectral gap function varying the the mixing term is illustrated to see the evolution of the dip near the anti-node.  

We note that electron pockets can also appear in the YRZ formulation, but have very different shapes \cite{leblanc}.

\section{Invariant pairings}
\label{app:pairing}
 
There are four combinations of invariant pairing terms of the $F$ and $G$:
\begin{eqnarray}
O^{A}_{\Delta}({i,j}) & =&\varepsilon^{\alpha\beta} (F_{i,\alpha}F_{j,\beta} + G_{i,\alpha}G_{j,\beta} ) \nonumber \\
O^{B}_{\Delta}({i,j})& =&\varepsilon^{\alpha\beta} (F_{i,\alpha}G_{j,\beta} + G_{i,\alpha}F_{j,\beta} ) \nonumber \\
O^{a}_{\Delta}({i,j}) & =&\varepsilon^{\alpha\beta} (-1)^{j_x+j_y}(F_{i,\alpha}F_{j,\beta} - G_{i,\alpha}G_{j,\beta} ) \nonumber \\
O^{b}_{\Delta}({i,j})& =&\varepsilon^{\alpha\beta} (-1)^{j_x+j_y}(G_{i,\alpha}F_{j,\beta} - F_{i,\alpha}G_{j,\beta} ) \label{ABtoFG}
\end{eqnarray}
In Table \ref{table2}, we illustrate the transformation of various pairing terms.
\begin{table}[t]
\begin{spacing}{2}
\centering
\begin{tabular}{|c|c|c|c|c|} \hline
 & $T_x$ & $R_{\pi/2}^{\rm dual}$ & $I_x^{\rm dual}$ & $\mathcal{T}$  \\
 \hline  \hline
$\varepsilon^{\alpha\beta} F_{\alpha}F_{\beta}$ & $\varepsilon^{\alpha\beta} G_{\alpha}G_{\beta}$ & $ \varepsilon^{\alpha\beta} G_{\alpha}G_{\beta}$ & $\varepsilon^{\alpha\beta} G_{\alpha}G_{\beta}$  &  -$\varepsilon^{\alpha\beta} F_{\beta}^\dagger F_{\alpha}^\dagger $ \\
\hline $\varepsilon^{\alpha\beta} G_{\alpha}G_{\beta}$ & $\varepsilon^{\alpha\beta} F_{\alpha}F_{\beta}$ & $ \varepsilon^{\alpha\beta} F_{\alpha}F_{\beta}$ & $\varepsilon^{\alpha\beta} F_{\alpha}F_{\beta}$  &  -$\varepsilon^{\alpha\beta} G_{\beta}^\dagger G_{\alpha}^\dagger $  \\ \hline
$\varepsilon^{\alpha\beta} F_{\alpha}G_{\beta}$ & $\varepsilon^{\alpha\beta} G_{\alpha}F_{\beta}$ & $ \varepsilon^{\alpha\beta} G_{\alpha}F_{\beta}$ & $\varepsilon^{\alpha\beta} G_{\alpha}F_{\beta}$  &  -$\varepsilon^{\alpha\beta} G_{\beta}^\dagger F_{\alpha}^\dagger $ \\
\hline $\varepsilon^{\alpha\beta} G_{\alpha}F_{\beta}$ & $\varepsilon^{\alpha\beta} F_{\alpha}G_{\beta}$ & $ \varepsilon^{\alpha\beta} F_{\alpha}G_{\beta}$ & $\varepsilon^{\alpha\beta} F_{\alpha}G_{\beta}$  &  -$\varepsilon^{\alpha\beta} F_{\beta}^\dagger G_{\alpha}^\dagger $  \\ \hline \hline
\end{tabular}
\end{spacing}
\caption{Transformations of the pairing terms. We suppress the lattice index($i,j$) before and after transformations. Note that the Time Reversal column $(\mathcal{T})$ contains $(-)$ term and the conjugate partner also have the $(-)$ sign. }
\label{table2}
\end{table}
The four pairings havean interesting exchange symmetry. 
Obviously $O^{A,B}_{\Delta}$ have even under the exchange operation. 
If we consider nearest neighbor sites,$(i,j)$, it is easy to show that $O^{b}_{\Delta}$ is even and $O^{a}_{\Delta}$ is odd under the exchange.
Therefore, for the $d_{x^2 -y^2}$ symmetry, the $O^{a}$ does not contribute to pairings.

The conversion between the two representations are as follows:
\begin{eqnarray}
O^{c}_{\Delta}({i,j}) & =&\varepsilon^{\alpha\beta} C_{i,\alpha}C_{j,\beta}  = \frac{1}{2} (O_{\Delta}^A + O_{\Delta}^B )(i,j) \nonumber \\
O^{d}_{\Delta}({i,j}) & =&\varepsilon^{\alpha\beta} D_{i,\alpha}D_{j,\beta} = \frac{(-1)^{\Delta x +\Delta y}}{2} (O_{\Delta}^A - O_{\Delta}^B)(i,j)\nonumber \\
O^{cd}_{\Delta}({i,j})& =&\varepsilon^{\alpha\beta} C_{i,\alpha}D_{j,\beta} = \frac{1 }{2}(O_{\Delta}^a + O_{\Delta}^b)(i,j) \nonumber \\
O^{dc}_{\Delta}({i,j})& =& \varepsilon^{\alpha\beta} D_{i,\alpha}C_{j,\beta} = \frac{(-1)^{\Delta x +\Delta y}}{2} (O_{\Delta}^a - O_{\Delta}^b)(i,j),
\end{eqnarray} 
where $\Delta x +\Delta y$ is coordinates' difference between two particles, for example,  zero for the $s$ wave and one for the $d$ wave.

\section{Pairing Instability} \label{Pairing Instability}
In this section, we introduce one way to achieve the $d$ wave instability from the gauge fluctuation. 
There could be many other channels to induce the $d$ wave channel such as ``conventional'' SDW fluctuation, so this section shows possibility of obtaining the desired pairings. 
   
\begin{table}[t]
\begin{spacing}{2}
\centering
\begin{tabular}{|c|c|c|c|c|} \hline
 & $T_x$ & $R_{\pi/2}^{\rm dual}$ & $I_x^{\rm dual}$ & $\mathcal{T}$  \\
 \hline  \hline
$\mathcal{B}$ & -$\mathcal{B}$& -$\mathcal{B}$ & +$\mathcal{B}$ &  -$\mathcal{B}$ \\
\hline 
$\mathcal{E}_x$ & -$\mathcal{E}_x$& -$\mathcal{E}_y$ & +$\mathcal{E}_x$ &  $\mathcal{E}_x$ \\ 
\hline
$\mathcal{E}_y$ & -$\mathcal{E}_y$& +$\mathcal{E}_x$ & -$\mathcal{E}_y$ &  $\mathcal{E}_y$ \\
\hline 
$\Psi$ &  $\tau^x \Psi$& $\tau^x \Psi$ &$\tau^x \Psi$&  $i\sigma^y (\Psi^{\dagger})^T$ \\
\hline \hline
\end{tabular}
\end{spacing}
\caption{Symmetry transformations of the $U(1)$ field strength of the CP$^1$ model, and 
of the fermion field $\Psi = (F \,\,G)^{T}$. }
\label{table3}
\end{table}

To constrain the Hamiltonian, let us consider symmetry transformations of the field strengths 
associated with the U(1) gauge field of the CP$^1$ model describing the $z_\alpha$ spinons
in Table \ref{table3} 
\begin{eqnarray}
&&\mathcal{B} = \Delta_x A_y - \Delta_y A_x~~,~~\mathcal{E}_x = \Delta_x A_\tau - \Delta_\tau A_x~~,~~\mathcal{E}_y = \Delta_y A_\tau - \Delta_\tau A_y \quad , \Psi =\begin{pmatrix} F \\ G  \end{pmatrix}.
\end{eqnarray}
Pauli matrix, $\tau(\sigma)$, is defined in the $(F,G)$ (spin) space.
The only invariant coupling up to the second order derivatives is \cite{pockets} 
\begin{eqnarray}
\label{eq:gamma}
\mathcal{S}_\gamma & =&  \gamma \int_{\tau,x}  {\bf \mathcal{E}}\cdot \Psi^{\dagger} \tau^y ( \nabla) \Psi \nonumber \\
& =& - i \gamma \int_{\tau,x} {\bf \mathcal{E}}\cdot (  F_\alpha^\dagger \nabla G_\alpha - G_\alpha^\dagger \nabla F_\alpha ).
\end{eqnarray}
It is interesting to note that this coupling is precisely the geometric phase coupling between the antiferromagnetic 
and valence bond solid (VBS) order parameters discussed recently in Ref.~\onlinecite{lsx}. The electric field is the spatial component
of the skyrmion currrent in the N\'eel state, and it couples here to a fermion operator which has the same quantum numbers as the spatial
gradient of the phase of VBS order; thus Eq.~(\ref{eq:gamma}) corresponds to the spatial terms in Eq.~(3.8)
in Ref.~\onlinecite{lsx}. Here we see that the electric field couples to a `dipole moment' in the fermions.

We can also look for a coupling between the magnetic field, $\mathcal{B}$, and the fermions.
There is no coupling up to the second order derivatives of fermionic fields. 
The main reason for the absence is that rotation and inversion transformations have opposite signs acting on the magnetic field. 
If we go beyond the second order derivative, we can find a coupling to the magnetic field such as 
\begin{eqnarray}
\mathcal{S}_{B} = \gamma_B \int_{\tau,x} \mathcal{B} \Psi^{\dagger} (\partial_x^2 - \partial_y^2)(\partial_x \partial_y) \tau^y \Psi.
\end{eqnarray}
This term is also one associated with the geometric phase between the antiferromagnetic and VBS orders, corresponds to the
temporal term in Eq.~(3.8)
in Ref.~\onlinecite{lsx}. 

The fluctuations of the gauge field are controlled by the action
\begin{eqnarray}
S_A &=&\frac{NT}{2} \sum_{\epsilon_n} \int \frac{d^2 k}{4 \pi^2} \left[   \Pi_{E} (k,\epsilon_n) |\mathcal{E}|^2  +  \Pi_{B}(k,\epsilon_n) |\mathcal{B}|^2 \right],
\label{sa}
\end{eqnarray}
where $ \Pi_{E} $ and $ \Pi_{B} $ are polarization functions from the matter fields. Because of the non-minimal coupling between
the electric and magnetic fields and the fermions, there is no screening, and these polarization functions are just constants at low momenta
and frequencies. Also, although the bosonic spinons do couple minimally to the electromagnetic field, they are gapped
and also yield only a constant contribution to the polarizations.

\begin{figure}
\includegraphics[width=2.0 in]{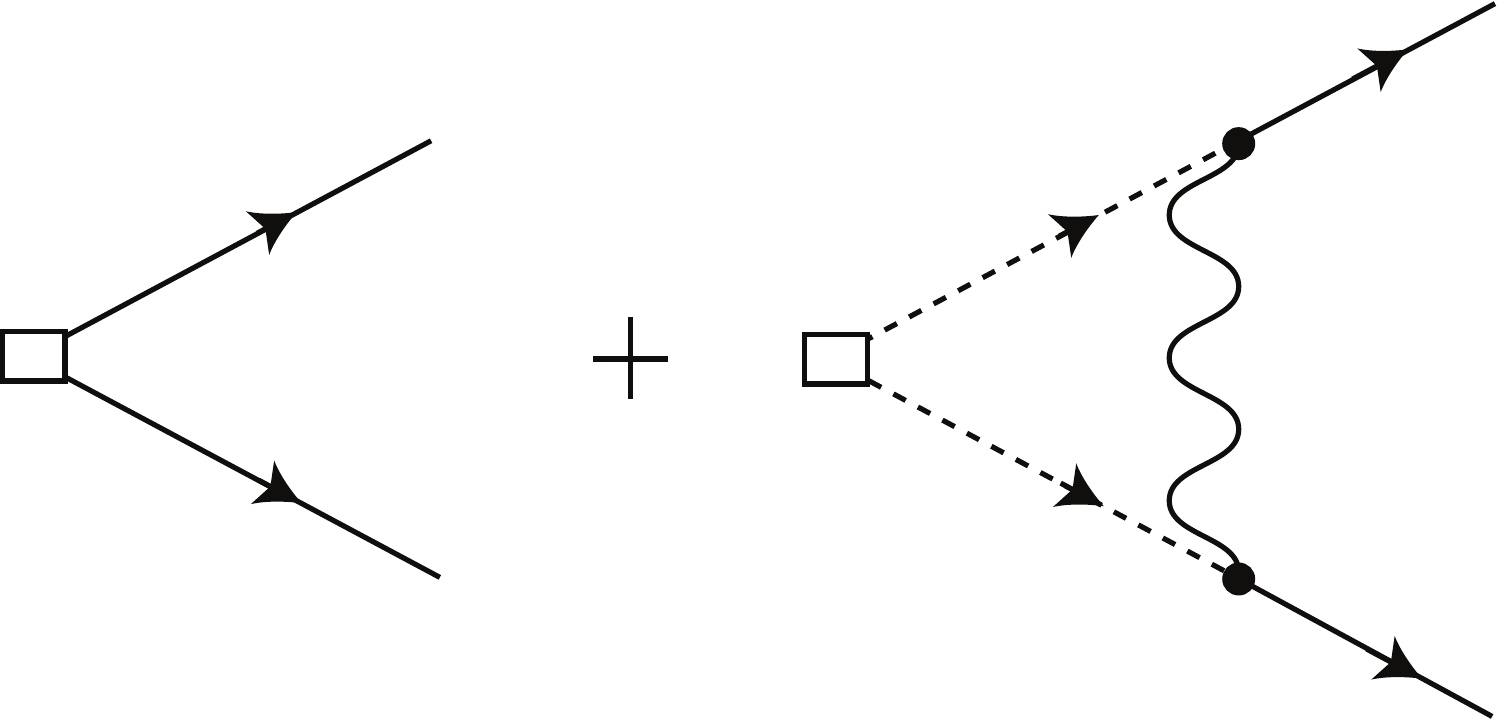}
\caption{ Vertex correction of a pairing channel. The wavy line is for the propagator of the electric fields, and the plain (dotted) line is for the $C(D)$ particle's. Note that the coupling with the electric field (filled dot) contains the momentum component. Here we represent the $C$ particle pairing vertex renormalization. }
\label{pairvertex}
\end{figure}
With the $C,D$ representation, the coupling term to the electric field becomes 
\begin{eqnarray}
\mathcal{S}_\gamma &=& -\gamma \int_{\omega, \Omega,k,q}  {\bf \mathcal{E}}(q,\Omega) \cdot {\bf k} \left[  D_\alpha^\dagger (k+q+Q,\omega+\Omega) C_\alpha(k,\omega) -C_\alpha^\dagger (k+q,\omega+\Omega) D_\alpha(k+Q,\omega)   \right]. \nonumber
\end{eqnarray}
It is manifest that $C$ and $D$ particles are only mixed with the finite momentum $Q$ difference. 

Let us consider pairing vertex
\begin{eqnarray}
V_{pairing} = \sum_{k} \delta^{c}(k) C^{\dagger}_{k, \uparrow} C^{\dagger}_{-k,\downarrow} +  \delta^{d}(k) D^{\dagger}_{k, \uparrow} D^{\dagger}_{-k,\downarrow} + h.c.
\end{eqnarray} 
To see the superconducting instability, we need to evaluate the vertex correction of the pairing channel such as the 
diagram in Fig.~\ref{pairvertex}. The presence of the $\lambda$ requires numerical evaluations. 
Instead of considering numerical calculations, let us turn off the mixing term, $\lambda$, and see which pairings are preferred with approximations. We will discuss about the non-zero mixing term later.
 
The renormalized pairing vertex of $C$ particles is 
\begin{eqnarray}
\delta^c (k)_{ren}  
& \sim&  \delta^c (k) -  \gamma^2 k_{F,d}^2 \delta^d (k+Q) \frac{\mathcal{N}_d}{\Pi_E} \int_{\varepsilon,\omega} \frac{ 1 }{\omega^2 + \varepsilon_{d}^2},
\end{eqnarray}
where $1/\Pi_E$ is the constant electric field propagator. 
As usual, we assume that the integration is dominant near Fermi surfaces and the $k^2$ becomes $k_F^2$.
Also we extract the gap function of $D$ particles out of the integration. The factor $\mathcal{N}_d$ is the density of states of $D$ particles.
Note that the minus sign in front of the second term is from the momentum dependence of the interaction and the relative sign of the gap functions.
Likewise, the $D$ particle pairing correction is
\begin{eqnarray}
\delta^d (k+Q)_{ren}  
& \sim& \delta^d (k+Q) - \gamma^2 k_{F,c}^2 \delta^c (k) \frac{\mathcal{N}_c}{\Pi_E} \int_{\varepsilon,\omega} \frac{ 1 }{\omega^2 + \varepsilon_{c}^2} \
\end{eqnarray}

In both equations, the last integrals show the usual BCS type logarithmic divergence. 
We can determine momentum dependence of the pairings with these equations. 
If we assume $s$ wave pairings, then the corrections become negative and the renormalized pairings become suppressed. 
On the other hand, $d$ wave pairings can change the sign of the integration and enhance the superconductivity. 
Such momentum dependence results from the momentum dependent vertex term in Eq.~(\ref{eq:gamma}) with a given relative pairing sign. 
In the gauge exchange, the momentum dependence plays the same role as spin-exchange in the usual $d$ wave BCS pairing. 

So far, we have fixed the relative sign between the two pairings by hand. 
Our calculation indicates possibility of $d$ wave pairings, but  the channel of the instability can vary with changing the relative pairing sign.
There could be fully gapped pairing with opposite  signs, $s^{\pm}$.

Evaluating the vertex corrections, we have assumed no mixing term, $\lambda$, at the lowest approximation. 
Now let us turn on the mixing term.
Then, the Fermi surfaces of the two particles start mixing and details of the Fermi surfaces change. 
Of course, $(C,D)$ pairings can be mixed by $\lambda$. 
But the mixing point is first gapped out and the Fermi surfaces become pockets.
So there is no significant pairing mixing by $\lambda$ and we can treat pairings separately. 
Details of the Fermi surface change, but we can argue that pairing channels remain intact at low energy.


\begin{thebibliography}{99}

\bibitem{boyer} M.~C.~Boyer, W.~D.~Wise, Kamalesh Chatterjee, Ming Yi, Takeshi Kondo, T. Takeuchi, 
H. Ikuta, and E.~W.~Hudson, Nature Physics {\bf 3}, 802 (2007).

\bibitem{dichot0}  Y.~Kohsaka, C.~Taylor, P.~Wahl, A.~Schmidt, Jhinhwan Lee, K.~Fujita, 
J.~W.~Alldredge, Jinho Lee, K.~McElroy, H.~Eisaki, S.~Uchida, D.-H.~Lee, J.~C.~Davis, 
Nature {\bf 454}, 1072 (2008). 

\bibitem{dichot1} Takeshi Kondo, Rustem Khasanov, Tsunehiro Takeuchi, J\"org Schmalian, and Adam Kaminski,
Nature {\bf 457}, 296 (2009).

\bibitem{dichot2} A. Pushp, C.~V.~Parker, A.~N.~Pasupathy, K. K. Gomes, S. Ono, J.~Wen, Z. Xu, G. Gu, and A.Yazdani, 
Science {\bf 324}, 1689 (2009).

\bibitem{dichot3} R.-H. He, K.Tanaka, S.-K. Mo, T. Sasagawa, M. Fujita, T. Adachi, N. Mannella, K.Yamada, Y.~Koike, Z. Hussain and Z.-X. Shen,
Nature Physics {\bf 5}, 119 (2008).

\bibitem{pockets} Y. ~Qi and S. ~Sachdev, Phys. Rev. B {\bf 81}, 115129 (2010).

\bibitem{ssss} S.~Sachdev, Phys. Rev. B {\bf 49}, 6770 (1994).

\bibitem{rkk1} R. K. Kaul, A. Kolezhuk, M. Levin, S. Sachdev, and T. Senthil, Phys. Rev. B  {\bf 75} , 235122 (2007).

\bibitem{rkk2} R. K. Kaul, Y. B. Kim, S. Sachdev, and T. Senthil, Nature Physics {\bf 4}, 28 (2008).

\bibitem{su2} S.~Sachdev, M. A. Metlitski, Y. Qi, and C. Xu, Phys. Rev. B {\bf 80}, 155129 (2009).

\bibitem{tsvelik1} M.~Khodas and A.~M.~Tsvelik, Phys. Rev. B {\bf 81}, 155102 (2010).

\bibitem{tsvelik2} M.~Khodas, H.-B.~Yang, J.~Rameau, P.~D.~Johnson, A.~M.~Tsvelik, 
and T.~M.~Rice, arXiv:1007.4837.

\bibitem{peter3} H.-B.~Yang, J.~D.~Ramaeu, Z.-H.~Pan, G.~D.~Gu, P.~D.~Johnson, 
R.~H.~Claus, D.~G.~Hinks, and T.~E.~Kidd,
 arXiv:1008.3121.

\bibitem{Meng} J.~Meng, G.~Liu, W.~Zhang, L.~Zhao, H.~Liu, X.~Jia, D.~Mu, S.~Liu, X.~Dong, W.~Lu, G.~Wang, Y.~Zhou, Y.~Zhu, X.~Wang, Z.~Xu, C.~Chen, X.~Zhou, Nature {\bf 462}, 335(2009)

\bibitem{Damascelli} D.~Fournier, G.~Levy, Y.~Pennec, J.~L.~McChesney, A.~Bostwick, E.~Rotenberg, R.~Liang, W.~N.~Hardy, D.~A.~Bonn, I.~S.~Elfimov and A.~Damascelli, Nature Physics {\bf 6}, 905 (2010)

\bibitem{andrey} T.~A.~Sedrakyan and  A.~V.~Chubukov, 
Phys. Rev. B {\bf 81}, 174536 (2010).

\bibitem{dmft1} M.~Civelli, M.~Capone, A.~Georges, K.~Haule, O.~Parcollet, T.~D.~Stanescu, 
and G. Kotliar, Phys. Rev. Lett. {\bf 100}, 046402 (2008).

\bibitem{dmft2} M.~Ferrero, P.~S.~Cornaglia, L.~De Leo, O.~Parcollet, G.~Kotliar, and A.~Georges,
Phys. Rev. B {\bf 80}, 064501 (2009).

\bibitem{sordi} G.~Sordi, K.~Haule, and A.-M.~S. Tremblay, Phys. Rev. Lett. {\bf 104}, 226402 (2010).

\bibitem{ffl1} T.~Senthil, S.~Sachdev, and M.~Vojta, Phys. Rev. Lett. {\bf 90}, 216403 (2003).

\bibitem{ffl2} T.~Senthil, M.~Vojta, and S.~Sachdev, Phys. Rev. B {\bf 69}, 035111 (2004).

\bibitem{paschen} J.~Custers,
P.~Gegenwart, C.~Geibel, F.~Steglich, P.~Coleman, 
and S. Paschen, 
Phys. Rev. Lett. {\bf 104}, 186402 (2010). 

\bibitem{osmt} V.~I.~Anisimov,  
I.~A.~Nekrasov, D.~E.~Kondakov, T.~M.~Rice, and M.~Sigrist, Eur. Phys. J. B {\bf 25}, 191
(2002); L.~De Leo, M.~Civelli, and G.~Kotliar, Phys. Rev. Lett. {\bf 101}, 256404 (2008).

\bibitem{vojta} M.~Vojta, J. Low. Temp. Phys. {\bf 161}, 203 (2010).

\bibitem{mv} O.~I.~Motrunich and A.~Vishwanath,
Phys.\ Rev.\  B {\bf 70}, 075104 (2004).

\bibitem{senthil} T.~Senthil, A.~Vishwanath, L.~Balents, S.~Sachdev, and
M.~P.~A.~Fisher, Science {\bf 303}, 1490 (2004).

\bibitem{rs2}  N. Read and S. Sachdev, Phys. Rev. Lett. {\bf 66}, 1773 (1991).

\bibitem{rsl} S. Sachdev and N. Read,
Int. J.
Mod. Phys. {\bf B5}, 219 (1991); arXiv:cond-mat/0402109.

\bibitem{wenholon} X.-G.~Wen, Phys. Rev. B {\bf 39}, 7223 (1989).

\bibitem{rs1} N.~Read and S.~Sachdev, Phys. Rev. Lett. {\bf 62}, 1694 (1989).

\bibitem{ss} B.~I.~Shraiman and E.~D.~Siggia, Phys. Rev. Lett. {\bf 61}, 467 (1988).

\bibitem{RW1} T.~C.~Ribeiro and X.-G.~Wen, Phys. Rev. B {\bf 74}, 155113 (2006).

\bibitem{RW2} Ying Ran and X.-G.~Wen, arXiv:cond-mat/0611034.

\bibitem{RW3} T.~C.~Ribeiro and X.-G.~Wen, Phys. Rev. B {\bf 77}, 144526 (2007).

\bibitem{ramshaw} B.~J.~Ramshaw, B.~Vignolle, J.~Day, Ruixing Liang, W.~N.~Hardy, Cyril Proust, 
and D.~A.~Bonn, arXiv:1004.0260.

\bibitem{gs} V.~Galitski and S.~Sachdev, Phys. Rev. B {\bf 79}, 134512 (2009).

\bibitem{superACL} E.~G. ~Moon and S.~Sachdev, Phys. Rev. B
  {\bf 80}, 035117 (2009). 
  
\bibitem{QCpairing} E.~G.~Moon and A.~Chubukov, arXiv:1005.0356


\bibitem{YRZ1}  Kai-Yu Yang, T. M. Rice and Fu-Chun Zhang, Phys. Rev. B
  {\bf 73}, 174501 (2006).

\bibitem{YRZ2} R.~M.~Konik, T.~M.~Rice, and A.~M.~Tsvelik,
Phys. Rev. Lett. {\bf 96}, 086407 (2006).
  
\bibitem{YRZ3} J.~P.~F.~Leblanc, J.~P.~Carbotte, and E.~J.~Nicol, Phys. Rev. B {\bf 80}, 060505 (2009).

\bibitem{YRZ4} E.~Schachinger and J.~P.~Carbotte,
Phys. Rev. B {\bf 81}, 214521 (2010).

\bibitem{YRZ5} A.~J.~H.~Borne, J.~P.~Carbotte, and E.~J.~Nicol,
Phys. Rev. B {\bf 82}, 024521 (2010).

\bibitem{WL1} X.-G.~Wen and P.~A.~Lee, Phys. Rev. Lett. {\bf 76}, 503 (1996).

\bibitem{WL2} P.~A.~Lee, N.~Nagaosa, Tai-Kai Ng, and X.-G.~Wen, Phys. Rev. B {\bf 57}, 6003 (1998).

\bibitem{schulz} H.~J.~Schulz, Phys. Rev. Lett. {\bf 65}, 2462 (1990); C.~Zhou and H.~J.~Schulz, Phys. Rev. B {\bf 52}, R11557 (1995).

\bibitem{leeholon}
P. A. Lee, Phys. Rev. Lett. {\bf 63}, 680 (1989).

\bibitem{shankar} R.~Shankar, Phys. Rev. Lett. {\bf 63}, 203 (1989).

\bibitem{ioffew} L.~B.~Ioffe and P.~B.~Wiegmann, Phys. Rev. Lett. {\bf 65}, 653 (1990).

\bibitem{css} A.~V.~Chubukov, T.~Senthil and S.~Sachdev, 
Phys. Rev. Lett. {\bf 72}, 2089 (1994).

\bibitem{leblanc} J.~P.~F.~LeBlanc and J.~P.~Carbotte, arXiv:1006.5034.

\bibitem{lsx} L.~Fu, S.~Sachdev, and C.~Xu, arXiv:1010.3745.
\end{thebibliography}
\end{document}